\documentclass[letter,oldversion]{aa}
\usepackage{graphicx}
\usepackage{dcolumn}
\newcommand{\ltsima}{$\; \buildrel < \over \sim \;$} 
\newcommand{\lesssim}{\lower.5ex\hbox{\ltsima}}   
\newcommand{\gtrsim}{\lower.7ex\hbox{$\;\stackrel{\textstyle>}{\sim}\;$}}   
\begin{document}

\title{Tidal dwarf galaxies as a test of fundamental physics}

\author{G. Gentile\inst{1}, B. Famaey\inst{2}, F. Combes\inst{3}, P. Kroupa\inst{4},
H. S. Zhao\inst{5}, and O. Tiret\inst{3}}

\offprints{G. Gentile}

\institute{University of New Mexico, Department of Physics and Astronomy, 800 Yale
Blvd NE, Albuquerque, NM 87131, USA \\ \email{ggentile@unm.edu}
\and
Institut d'Astronomie et d'Astrophysique, Universit\'e Libre de Bruxelles, CP 226,
Boulevard du Triomphe, B-1050 Bruxelles, Belgium
\and 
LERMA, Observatoire de Paris, 61 Av. de l'Observatoire, F-75014 Paris, France
\and
Argelander-Institut f\"ur Astronomie (AIfA), Universit\"at Bonn, Auf dem H\"ugel 71, D-53121 Bonn, Germany 
\and 
SUPA, School of Physics and Astronomy, University of St. Andrews, KY16 9SS, Fife, UK
}

\date{\today}

\abstract{Within the cold dark matter (CDM) framework tidal dwarf galaxies (TDGs) 
cannot contain dark matter, so the recent results 
by Bournaud et al. (2007) that 3 rotating TDGs do show significant evidence for being dark matter dominated 
is inconsistent with the current concordance cosmological theory unless yet another dark matter component is postulated.
We confirm that the TDG rotation curves are consistent with Newtonian dynamics only if
either an additional dark matter component is postulated, or if
all 3 TDGs happen to be viewed nearly edge-on, which is unlikely given the geometry of the tidal debris.
We also find that the observed rotation curves are very naturally explained without any free parameters 
within the modified Newtonian dynamics (MOND) framework if inclinations are adopted as derived 
by Bournaud et al.  We explore different inclination 
angles and two different assumptions about the external field effect. The results do not change significantly, and we 
conclude therefore that Newtonian dynamics has severe problems while
MOND does exceedingly well in
explaining the observed rotation curves of the 3 TDGs studied by Bournaud et al.
}


   \keywords{Gravitation -- dark matter -- 
Galaxies: dwarf -- Galaxies: kinematics and dynamics}

\authorrunning{Gentile et al.}


\maketitle

\section{Introduction}

Dwarf irregular galaxies have the tendency to form during galaxy interactions 
within the extended tidal tails,
and are therefore dubbed Tidal Dwarf Galaxies (TDG, Mirabel, Dottori \& Lutz 1992; Duc et al 2000,
Braine et al. 2000).
Tidal tails transport the angular momentum and energy away from the merging sub-systems which build up the larger 
galaxies we see today, and gravitational instabilities within the tidal tails lead to local collapse and star formation. 
Observations of local interacting galaxies 
have shown the formation of TDGs to be quite common, in some cases dozens 
of condensations are seen in the tidal 
tails (Weilbacher et al. 2000), with a few ones having a mass typical of dwarf galaxies (Braine et al 2000). Based on a sample of 6 local interacting galaxies, 
Delgado-Donate et al. (2003) estimate that at most a few long-lived TDGs forms per merger.
Monreal-Ibero et al (2007) studied in detail the stability of external star forming regions in ULIRGs, and concluded that
they are good TDG candidates.
The relation in terms of evolution beween TDGs and the other kinds of dwarf galaxies is
presently a matter of debate (see, e.g., Duc \& Mirabel 1998, Okazaki et Taniguchi 2000, Bournaud \& Duc 2006,
Metz \& Kroupa 2007). In particular, Okazaki \& Taniguchi (2000) claim 
that conservative assumptions about TDG production within a hierarchical 
CDM structure formation framework imply that all dE galaxies may be TDGs 
leaving little or no room for traditional dark-matter filled dwarf galaxies
(but see Bournaud \& Duc 2006). On the other hand, Skillman \& Bender (1995)
cast some doubt on the idea that all dwarf ellipticals originate from dwarf irregulars 
after their gas has been blown away.

In any case, in the Cold Dark Matter (CDM) scenario, TDGs should be nearly 
devoid of dark matter (Barnes \& Hernquist 1992), contrary to other kinds of dwarf galaxies.

In a recent paper, Bournaud et al. (2007) (hereafter B07) analyse the rotation curves of 3 TDGs 
belonging to the NGC 5291 system. They find evidence for a mass discrepancy that is 
unexpected within the CDM framework, and they put forward the hypothesis of baryonic dark matter
to explain the observations. 
Here we investigate the possibility that the observations can
be explained within the framework of MOND (Modified Newtonian Dynamics, Milgrom 1983), without the need
for baryonic dark matter.

In general, disk galaxies' rotation curves do not decrease in the outer parts  
as would be expected from the visible matter
distribution. In the standard picture, this is explained by dark matter halos around galaxies. 
However, MOND is an alternative explanation where our understanding of gravity (or inertia) 
is changed, 
rather than our 
understanding of the matter content of the Universe. Milgrom (1983) postulated that for gravitational accelerations 
below $a_0 \approx 10^{-8} {\rm cm} \, {\rm s}^{-2}$ the effective gravitational attraction approaches $(g_N a_0)^{1/2}$ 
where $g_N$  
is the usual Newtonian gravitational field. MOND leads 
to remarkable fits of galactic kinematics over 5 decades in mass (Sanders \& McGaugh 2002) ranging from tiny 
dwarfs (e.g., Gentile et al. 2007) through our own Milky Way (Famaey \& Binney 2005; Famaey, Bruneton \& Zhao 2007) 
and early-type spirals (e.g., Sanders \& Noordermeer 2007), to massive ellipticals (Milgrom \& Sanders 2003), without 
resorting to galactic dark matter.
However, possible problems for MOND on galactic scales include 
the kinematics of Local Group dwarf spheroidals ({\L}okas, Mamon \& 
Prada 2006) and galaxy
merging timescales (Nipoti et al. 2007).

TDGs are thus objects on galactic scales on which to test the MOND paradigm\footnote {A preliminary 
fit just posted on the arXiv by Milgrom (2007) has shown that MOND has the ability to 
explain the dynamics without resorting to disk dark matter.}.
Here we take into account uncertainties on the external field 
effect (EFE) and the galaxies' inclinations to show that the MOND paradigm is very likely to correctly 
explain the kinematics of the TDGs.

\section{Data}

We use the HI rotation curves recently published by B07.
They performed HI observations of the NGC 5291 system, using the VLA 
(Very Large Array) in its
BnA and CnB configuration, yielding a spatial resolution of $\sim$ 2.2 $\times$ 1.6 kpc.
They discuss the ring structure composed of collisional debris that surrounds NGC 5291, and they detect eight
clumps where velocity gradients can be observed. Only three of these, however,
are resolved enough to allow a more detailed investigation of their kinematics.

The rotation curves of these three TDGs (NGC5291N, NGC5291S,
and NGC5291SW) were derived by B07 using
an envelope-tracing method (Sancisi \& Allen 1979; Sofue 1996; Gentile et al. 2004),
which gives reliable velocity estimates for poorly resolved and/or highly inclined
disk galaxies. For the purpose of our analysis,
we folded the two sides of the rotation curve $V_{\rm rot}(r)$ and the baryonic contribution
$V_{\rm bar} (r)$, to 
obtain azimuthally averaged $V_{\rm rot}(r)$ and $V_{\rm bar} (r)$ curves.
B07 use an inclination $i$ of 45$^{\circ}$ for the three TDGs, based on the fact that 
their model reproduces successfully the morphology of the system with $i=45^{\circ}$,
and that the rotation axis of the clumps differs from that of the ring by less
than 15$^{\circ}$. 
The total baryonic masses, as derived by B07, are: $9^{+1.0}_{-0.7} 
\times 10^8$ M$_{\odot}$ for
NGC 5291N, $9.3^{+1.1}_{-0.9} \times 10^8$ M$_{\odot}$ for
NGC 5291S, and $5 \pm 1.5 \times 10^8$ M$_{\odot}$ for
NGC 5291SW. 

\section{Fitting procedure}
\label{procedure}

To make a rigorous fit of these galaxies in MOND, two important issues to take into account are their inclination and the EFE of MOND (e.g., Famaey et al. 2007; Wu et al. 2007, Angus \& McGaugh 2007). 
The rotation curves were thus fitted within the MOND framework with five different
hypotheses:

$\bullet$ First, we simply applied the MOND prescription to the rotation curve. This approach
has no {\it explicit} free parameter: we used $i$, the distance and the stellar mass-to-light
($M/L$) ratio as derived by B07. The $M/L$ ratio is not a crucial assumption
since the stellar mass, derived from stellar population synthesis models, is 
several times lower than the total baryonic mass. We used $a_0 = 1.2 \times 10^{-8}$ cm s$^{-2}$ (Begeman,
Broeils \& Sanders 1991).

$\bullet$ Second, we left $i$ as a free parameter. Indeed $i$ is an important parameter influencing the 
derived dynamical mass, and the 
inclination determined by B07 is only an estimate. The inclination we used is however constant with radius: i.e.,
we ignored any possible warps,
which are   
taken into account in the error-bars.

$\bullet$ Third, we added a first estimate for the EFE, while $i$ was left as a free parameter. In modified gravity (Bekenstein \& Milgrom 1984), if one considers, as a first approximation, that a galaxy free-falls
with a ``uniform" acceleration in an external linear potential, the internal potential becomes polarized Keplerian (Milgrom 1986, Zhao \& Tian 2006, Zhao \& Famaey 2006). Here, to simplify the treatment of the EFE, we follow the less rigorous approach of Famaey et al. (2007) in which  the polarization in the direction of the external field (EF) as well as a term proportional to the gradient of $\mu$ are neglected.  We define the internal acceleration as 
$a_{\rm int} = V^2_{\rm obs} / r$, where $V_{\rm obs}$ is 
the observed rotation velocity and $r$ is the galactocentric 
radius in the TDGs. Then, following the MOND hypothesis: $a_{\rm int} = 
g_{\rm N} / \mu(x)$, where $g_{\rm N} = V^2_{\rm bar} / r$
is the Newtonian acceleration, $\mu(x) = x / (1+x)$ (Famaey \& Binney 2005) and 
$x = (a_{\rm int} + g_{\rm ext}) / a_0$
(instead of the usual $x = a_{\rm int} / a_0$), where $g_{\rm ext}$
is the EF.
We roughly estimate $g_{\rm ext}$ due to NGC 5291 as follows: 
from the total HI profile $\Delta V_{20}$ (Malphrus et al. 1997),
and assuming that the rotation curve of NGC 5291 stays flat
out to the distances of the TDGs, 
we find $g_{\rm ext} \lesssim (0.5 \Delta V_{20} / sin(i))^2~/R_{\rm G}$,
where $R_{\rm G}$ is the projected distance of the TDGs from 
the centre of NGC 5291. This gives $g_{\rm ext} \lesssim
0.3 a_0$. We then assume $g_{\rm ext}$ to have a typical value
of 0.2$a_0$. Hence, the galaxies are mostly dominated by $a_{\rm int}$ until the last data point, 
so the various approaches to take the EF into account are slightly degenerate, which we illustrate with 
$x={\rm max}(a_{\rm int}/a_0,g_{\rm ext}/a_0)$ as a
second parametrization. Our two choices of $x$ for the EFE bracket the range
of possible effects of the EF on the rotation curve, the first one giving a larger EFE than the second.
A more rigourous treatment of the effect of EF on rotation curves has already been done
in Wu et al. (2007). It was found that the internal iso-potential contours are twisted
due to the general orientations of the EF,
but there are typically less than 10\% difference among
different orientations (hence the modulus of vector sums) of the EF and internal
field.  Rotation curves would become normal baryon-only Newtonian RCs if the system
accelerates faster and faster in a stronger and stronger EF.

$\bullet$ Fourth, we fitted the rotation curves with Newtonian gravity and no dark matter, as would
be expected in CDM, still keeping $i$ as a free parameter.

$\bullet$ Finally, we fixed $i$ at the most favourable inclination for the 
previous fits, and we derived the baryonic distribution required to fit the rotation curves
in Newtonian gravity.

\begin{figure*}
\hspace{2.2cm}
{\includegraphics[width=140mm]{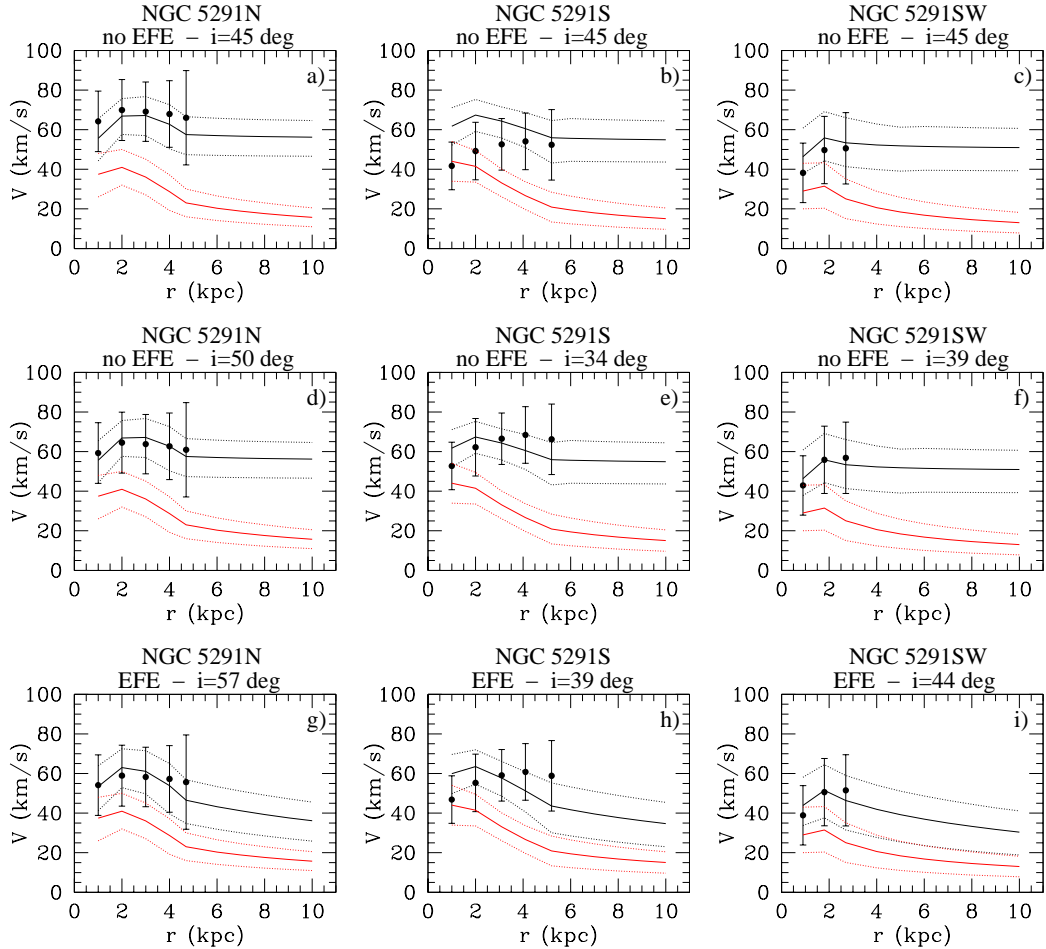}}
\caption{Rotation curve data (full circles)
of the 3 tidal dwarf galaxies (Bournaud et al. 2007).
The lower (red) curves are the Newtonian contribution $V_{\rm bar}$ of the 
baryons (and its uncertainty, indicated as dotted lines).
The upper (black) curves are the MOND prediction and its uncertainty
(dotted lines).
The top panels have as an implicit assumption (following Bournaud et al.)
an inclination angle of 45 degrees.
In the middle panels the inclination is a free parameter,
and the bottom panels show the fits made with the first estimate for 
the external field effect (EFE) (Section \ref{procedure}).
}\label{fits}
\end{figure*}

\section{Results}
\label{section_results}

\begin{figure}
\hspace{0.4cm}
{\includegraphics[width=80mm, angle=0]{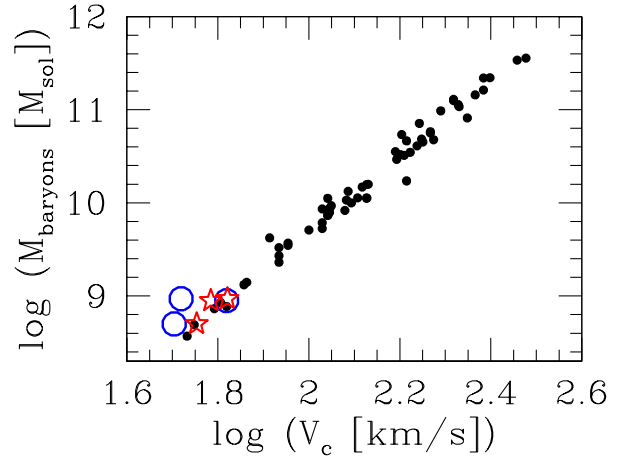}}
\caption{
Baryonic Tully-Fisher relation (baryonic mass vs. circular velocity). The
small full circles are the disk galaxy data from McGaugh (2005).
The 3 TDGs studied here are shown as blue empty circles ($i=45^{\circ}$)
and red stars ($i$ as a free parameter).
}
\label{btf}
\end{figure}

In Fig. \ref{fits} we show the results
of the fits. Overall MOND gives good fits to the 
rotation curves of these systems. In particular, 
Fig. \ref{fits} (panels a - c) shows that with no explicit free parameter, MOND can explain well the 
observed kinematics. As could be expected a priori, leaving 
$i$ as a free parameter improves the quality of the fits
(see Fig. \ref{fits}, panels d - f). The best-fit
values are not far from the first assumption of $i=45^{\circ}$: we find
$i=50^{+10}_{-7}$ degrees for NGC 5291N,
$i=34 \pm 4 $ degrees for NGC 5291S,
and $i=39^{+11}_{-7}$ degrees for NGC 5291SW: their average and scatter 
is similar to B07.
It is interesting to note that all three TDGs fall on the 
baryonic Tully-Fisher (BTF) relation (McGaugh 2005): the agreement
is perfect for $i$ as a free parameter and only slightly worse for $i=45^{\circ}$
(see Fig. \ref{btf}).
TDGs add
to a long list of classes of galactic objects that satisfy the BTF
relation.  It is puzzling that an empirical relation works so well for galaxies
of a wide range of formation histories unless it is the direct prediction of a
realistic law of gravity relating to the instantaneous baryon
distribution, as MOND is.

The fits with the two EFE estimates (see Fig. \ref{fits}, panels g - i, for
the first estimate)
give good fits, whose best-fit parameters are displayed in Table \ref{bestfit}.
The first EFE estimate gives slightly worse fits than the second;  
we thus conclude that the EFE due to NGC 5291 in the 3 TDGs cannot be too large:
it has to be smaller than $\sim (0.1-0.2) a_0$.

We realise that the EFE is a function of distance $R_{\rm G}$ from the host galaxy NGC 5291.
The 3 TDGs are at projected distances ranging from 58 to 75 kpc; given the small range of 
projected distances, and since the 3D distance is unknown, the EFE estimates are made using their
average projected distance, 65 kpc, instead of their individual projected distances, 
which results in errors on the EFE $\lesssim 15\%$, much
smaller than the overall uncertainties on the actual EFE value. 
We ran models of the EFE with variations of $\sim 15\%$, and found very
similar results.
Similarly, the difference
between the EF effects on the two sides of the TDG rotation curves,
is smaller than the uncertainty on its value, so we ignored it and worked on the 
azimuthally averaged rotation curve.

In Fig. \ref{fits} the curves with EFE
show a Keplerian fall-off just outside the present data, a signal which is
falsifiable if future observations could extend the kinematics of these TDGs.  This is
not due to tidal truncation, but to the EFE.
A rotating disk
inclined with an angle $\theta_{EF}$ with respect to the EF will generally precess
around the axis of the EF (Wu et al. 2007, Famaey et al. 2007). 
The precession means that these TDG disks in MOND are not expected
to be exactly aligned with the same inclinations, justifying our fits where 
$i$ is left as a free parameter.  
We note that continued accretion from the tidal matter and torquing of the TDGs may, however, 
affect the shape of the rotation curve at large radii, so the solid lines 
at large radii only show the case for unperturbed satellites.

\begin{figure}
\hspace{0.4cm}
{\includegraphics[width=83mm, angle=0]{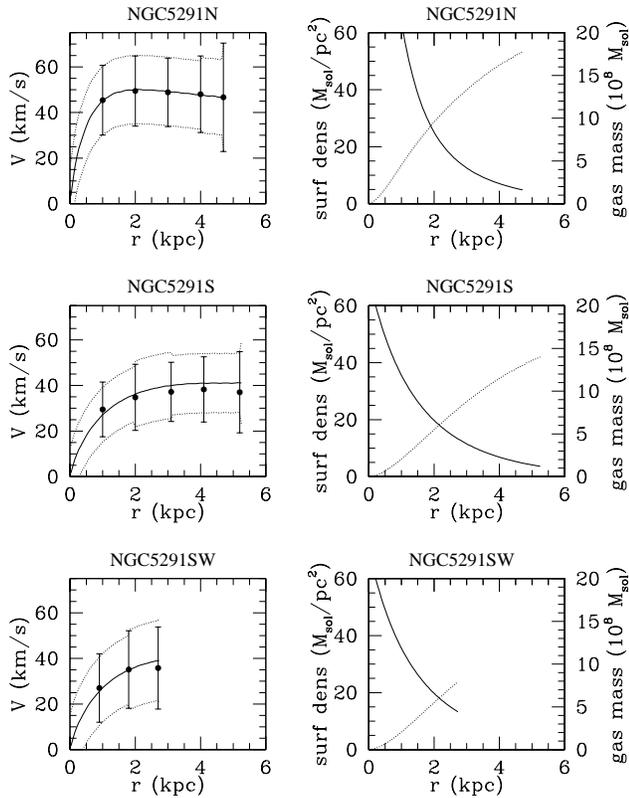}}
\caption{
Left: rotation curves with $i=90^{\circ}$, with the
baryonic distribution that best fits them in Newtonian dynamics. 
Symbols and curves are like in Fig. \ref{fits}.
Right: the corresponding 
surface density distribution (solid line) and integrated mass (dotted line).
}
\label{baryons}
\end{figure}

We note that large uncertainties still exist in
the mass distributions and inclinations of these TDGs,
and thus in their dynamics. 
The scenario proposed by B07 within the CDM paradigm leaves some liberty as to the exact 
geometry and encounter.
In particular, the inclination of the TDG on the sky planes
could be more edge-on,
and the inferred enclosed mass required to explain the observed
velocities smaller, decreasing the need for dark matter
or modified gravity.
The overall quality of Newtonian, purely baryonic fits with $i=90^{\circ}$ 
and with the baryonic distribution taken from B07 (not shown in Fig. \ref{fits})
is worse than the MOND 
fits with a free inclination, but it is only slightly worse than the 
MOND fits with $i=45^{\circ}$: in NGC5291N the MOND fit is much better, in 
NGC5291S it is a bit worse, and in NGC5291 the two fits are equivalent.
However, given the geometry of the NGC 5291 system,
it seems unlikely (but possible) that all 3 TDGs studied have 
such high inclinations. 
Also, the radial distribution of baryons, essentially the
HI gas, is also poorly determined, and can be chosen
to fit the observed rotation curves in Newtonian gravity, instead of being selected
a priori.
We have made such a fit (Fig. \ref{baryons}), by selecting exponential profiles for
the surface density of the interstellar gas
distributed in a disk. The enclosed
gas mass within each radius is also plotted, together with the
gas surface density, compatible with the observations.
The deduced rotation curves are consistent with the data,
within the error bars.
Let us note that the fit is still rough, no effort was made to
fit more components (such as the stellar component and the molecular gas)
in addition to the HI gas.

Given the
(relatively) poor quality of the rotation curves, 
a more rigorous study of the EFE and of the possible warps is difficult,
since they are not 
well constrained observationally. The uncertainty on the EFE is unlikely to
significantly affect our conclusions given that the EFE and the no-EFE results
do not differ much.

\begin{table}
\caption{Best-fit values of the 5 fits for each TDG. $i$ is the inclination,
and the EFE estimates refer to Section \ref{procedure}. The 1-$\sigma$ uncertainties
in the fitted inclinations are based on the $\chi^2$ statistics.}
\begin{tabular}{cccc}
\hline
TDG name    &     $\chi^2$     &     $i$                 &  note   \\
\hline
NGC 5291N   &      0.60        &   45 deg                & no EFE     \\       
            &      0.16        &   50$^{+10}_{-7}$ deg   & no EFE    \\
            &      0.30        &   57$^{+17}_{-9}$ deg   & first EFE estimate  \\
            &      0.17        &   50$^{+10}_{-8}$ deg   & second EFE estimate \\ 
            &      3.60        &   90$^{+0}_{-16}$ deg   & baryons + Newt. gravity \\ 
\hline
NGC5291S    &      5.42        &   45 deg                & no EFE    \\       
            &      1.36        &   34$\pm$4 deg          & no EFE    \\
            &      2.70        &   39$^{+6}_{-5} $ deg   & first EFE estimate  \\
            &      1.70        &   35$\pm$4 deg          & second EFE estimate \\ 
            &      3.25        &   90$^{+0}_{-30}$ deg  & baryons + Newt. gravity \\ 
\hline
NGC5291SW   &      0.45        &   45 deg                & no EFE    \\       
            &      0.09        &   39$^{+11}_{-7}$ deg   & no EFE    \\
            &      0.19        &   44$^{+16}_{-9}$ deg   & first EFE estimate  \\
            &      0.09        &   39$^{+11}_{-7}$ deg   & second EFE estimate \\ 
            &      0.42        &   90$^{+0}_{-33}$ deg   & baryons + Newt. gravity \\ 
\hline
\label{bestfit}
\end{tabular}
\end{table}

\section{Conclusions}

We analysed the rotation curves of 3 TDGs 
presented by Bournaud et al. (2007)
and we found that MOND explains their kinematics
very naturally. Within the MOND framework, no additional unseen
matter is required in these 3 TDGs.
The dark matter required in Newtonian dynamics
implies a large amount of unseen baryonic
matter, which is very unconventional in the CDM context.

We explored different possibilities for the inclination angles
and the external field effect (EFE) due to NGC 5291. The results
do not change significantly: obviously the fits are improved with
one extra free parameter (the inclination), and we find  
the external field to be smaller than $\sim (0.1-0.2) a_0$,
within the a priori expectations from the kinematics of NGC 5291.
Therefore, MOND can fit the kinematics of the 3 TDGs presented
by Bournaud et al. (2007) for inclinations similar to those that 
they estimated and for realistic values of the external field.

We also found that in the (rather unlikely) case where all 3 TDGs are seen edge-on,
Newtonian fits with no dark matter are possible but slightly worse than the MOND fits
with $i=45^{\circ}$, and not as good as the MOND fits with 
$i$ as a free parameter.
This Letter has thus shown that TDGs pose severe challenges
to the current standard cosmological theory and that very recent observations
of rotating TDGs are consistent with a non-Newtonian interpretation. TDGs thus hold important
clues on fundamental physics and therewith require further research.

\acknowledgements{We thank S. McGaugh and G. Angus for useful suggestions, 
F. Bournaud for providing the kinematical data, and the anonymous
referee for useful suggestions.}



\begin{thebibliography}{}

\bibitem[Angus \& McGaugh(2007)]{2007arXiv0704.0381A} Angus, G.~W., \& McGaugh, S.~S.\ 2007, 
MNRAS submitted (arXiv:0704.0381)
\bibitem[Barnes \& Hernquist(1992)]{1992Natur.360..715B} Barnes, J.~E., \& Hernquist, L.\ 1992, 
Nature, 360, 715 
\bibitem[Begeman et al.(1991)]{1991MNRAS.249..523B} Begeman, K.~G., 
Broeils, A.~H., \& Sanders, R.~H.\ 1991, MNRAS, 249, 523 
\bibitem[Bournaud \& Duc(2006)]{2006A&A...456..481B} Bournaud, F., 
\& Duc, P.-A.\ 2006, A\&A, 456, 481 
\bibitem[Bournaud et al.(2007)]{2007arXiv0705.1356B} Bournaud, F., 
et al.\ 2007, Science, 316, 1166 (arXiv:0705.1356)
\bibitem[Braine et al.(2000)]{2000Natur.403..867B} Braine, J., Lisenfeld, 
U., Duc, P.-A., 
\& Leon, S.\ 2000, Nature, 403, 867 
\bibitem[Calura \& Matteucci(2004)]{2004MNRAS.350..351C} Calura, F., \& Matteucci, F.\ 2004, MNRAS, 350, 351 
\bibitem[Delgado-Donate et al.(2003)]{2003A&A...402..921D} Delgado-Donate, E.~J., 
Mu{\~n}oz-Tu{\~n}{\'o}n, C., Deeg, H.~J., \& Iglesias-P{\'a}ramo, J.\ 2003, A\&A, 402, 921
\bibitem[Duc \& Mirabel(1998)]{1998A&A...333..813D} Duc, P.-A., \& Mirabel, I.~F.\ 1998, A\&A, 333, 813 
\bibitem[Duc et al.(2000)]{2000AJ....120.1238D} Duc, P.-A., Brinks, E., Springel, V., Pichardo, B., 
Weilbacher, P., \& Mirabel, I.~F.\ 2000, AJ, 120, 1238 
\bibitem[Famaey \& Binney(2005)]{2005MNRAS.363..603F} Famaey, B., \& 
Binney, J.\ 2005, MNRAS, 363, 603 
\bibitem[Famaey et al.(2007)]{2007MNRAS.377L..79F} Famaey, B., Bruneton, 
J.-P., \& Zhao, H.\ 2007, MNRAS, 377, L79 
\bibitem[Gentile et al.(2004)]{2004MNRAS.351..903G} Gentile, G., Salucci, 
P., Klein, U., Vergani, D., \& Kalberla, P.\ 2004, MNRAS, 351, 903 
\bibitem[Gentile et al.(2007)]{2007MNRAS.375..199G} Gentile, G., Salucci, 
P., Klein, U., \& Granato, G.~L.\ 2007, MNRAS, 375, 199 
\bibitem[{\L}okas et al.(2006)]{2006EAS....20..113L} {\L}okas, E., Mamon, G., \& Prada, F.\ 2006, EAS Publications Series, 20, 113 
\bibitem[McGaugh(2005)]{2005ApJ...632..859M} McGaugh, S.~S.\ 2005, ApJ, 632, 859 
\bibitem[Metz \& Kroupa(2007)]{2007MNRAS.376..387M} Metz, M., \& Kroupa,
P.\ 2007, MNRAS, 376, 387
\bibitem[Milgrom(1983)]{1983ApJ...270..365M} Milgrom, M.\ 1983, ApJ, 270, 365 
\bibitem[Milgrom(1986)]{1986ApJ...302..617M} Milgrom, M.\ 1986, ApJ, 302, 617 
\bibitem[Milgrom \& Sanders(2003)]{2003ApJ...599L..25M} Milgrom, M., \& 
Sanders, R.~H.\ 2003, ApJ, 599, L25 
\bibitem[Milgrom(2007)]{2007arXiv0706.0875M} Milgrom, M.\ 2007, preprint (arXiv:0706.0875) 
\bibitem[Mirabel et al.(1992)]{1992A&A...256L..19M} Mirabel, I.~F.,
Dottori, H., \& Lutz, D.\ 1992, A\&A, 256, L19
\bibitem[Monreal-Ibero et al.(2007)]{2007arXiv0706.1145M} Monreal-Ibero, 
A., Colina, L., Arribas, S., \& Garcia-Marin, M.\ 2007, A\&A in press (arXiv:0706.1145) 
\bibitem[Nipoti et al.(2007)]{2007arXiv0705.4633N} Nipoti, C., Londrillo, P., \& Ciotti, L.\ 2007, ArXiv e-prints, 705, arXiv:0705.4633 
\bibitem[Okazaki \& Taniguchi(2000)]{2000ApJ...543..149O} Okazaki, T., \& Taniguchi, 
Y.\ 2000, ApJ, 543, 149
\bibitem[Sancisi \& Allen(1979)]{1979A&A....74...73S} Sancisi, R., \& 
Allen, R.~J.\ 1979, A\&A, 74, 73 
\bibitem[Sanders \& McGaugh(2002)]{2002ARA&A..40..263S} Sanders, R.~H., \& 
McGaugh, S.~S.\ 2002, ARA\&A, 40, 263 
\bibitem[Sanders \& Noordermeer(2007)]{2007astro.ph..3352S} Sanders, R.~H., 
\& Noordermeer, E.\ 2007, MNRAS in press (astro-ph/0703352)
\bibitem[Skillman \& Bender(1995)]{1995RMxAC...3...25S} Skillman, E.~D., \& Bender, R.\ 1995, Revista Mexicana de Astronomia y Astrofisica Conference Series, 3, 25
\bibitem[Sofue(1996)]{1996ApJ...458..120S} Sofue, Y.\ 1996, ApJ, 458, 120 
\bibitem[Weilbacher et al.(2000)]{2000A&A...358..819W} Weilbacher, P.~M., Duc, P.-A., 
Fritze v.~Alvensleben, U., Martin, P., \& Fricke, K.~J.\ 2000, A\&A, 358, 819
\bibitem[Wu et al. (2007)]{wu07} Wu, X., Zhao, H. S., Famaey, B.,  Gentile, G. , Tiret, O., Combes, F., Angus, G. W., Robin, A., 2007, arXiv:0706.3703
\bibitem[Zhao \& Famaey(2006)]{2006ApJ...638L...9Z} Zhao, H.~S., \& Famaey, 
B.\ 2006, ApJ, 638, L9 
\bibitem[Zhao \& Tian(2006)]{2006A&A...450.1005Z} Zhao, H. S., \& Tian, L.\ 
2006, A\&A, 450, 1005 


\end{thebibliography}
\end{document}